%% file: eprint.tex
\documentclass[10pt]{article}
\usepackage{graphicx}
\usepackage{url}
\usepackage{lmodern, amsmath, amssymb, xcolor}

\usepackage[font=small,labelfont=bf]{caption}

\def\Title#1{\begin{center} {\Large #1 } \end{center}}
\def\Author#1{\begin{center}{ \sc #1} \end{center}}
\def\Address#1{\begin{center}{ \it #1} \end{center}}

\newcommand\pubblock{\rightline{\begin{tabular}{l} Proceedings of the Fifth Annual LHCP\\ \pubnumber\\
         \pubdate  \end{tabular}}}

\newenvironment{Abstract}{\begin{quotation} \begin{center} 
             \large ABSTRACT \end{center}\bigskip 
      \begin{center}\begin{large}}{\end{large}\end{center} \end{quotation}}

\newenvironment{Presented}{\begin{quotation} \begin{center} 
             PRESENTED AT\end{center}\bigskip 
      \begin{center}\begin{large}}{\end{large}\end{center} \end{quotation}}

\def\Acknowledgements{\bigskip  \bigskip \begin{center} \begin{large}
             \bf ACKNOWLEDGEMENTS \end{large}\end{center}}

\input econfmacros.tex

\usepackage{color}

 \newcommand\pubnumber{ }

\newcommand\pubdate{\today}

\def\affiliation{
Institute of High Energy Physics, Chinese Academy of Sciences, Beijing 100049, China}

\newcommand*{\tmp}[4]{%
	{#4%
	\ifx\empty#3\empty\ifx\empty#1\empty\else^{#1}\fi\else^{#1(#3)}\fi%
	\ifx\empty#2\empty\else_{#2}\fi}%
}

\newcommand*{\qq }[3]{\tmp{#1}{#2}{#3}{O}}
\renewcommand{\hc}[1]{{#1}}
\newcommand{\FDF}{(\varphi^\dagger i\!\!\overleftrightarrow{D}_\mu\varphi)}
\newcommand{\FDFI}{(\varphi^\dagger i\!\!\overleftrightarrow{D}^I_\mu\varphi)}

\begin{document}

\large
\begin{titlepage}
\pubblock

\vfill
\Title{Theoretical aspects of the study of top quark properties}
\vfill

\Author{ Cen Zhang  }
\Address{\affiliation}
\vfill
\begin{Abstract}

We review some recent theoretical progresses towards the determination of the
top-quark couplings beyond the standard model.  We briefly introduce the global
effective field theory approach to the top-quark production and decay
processes, and discuss the most useful observables to constrain the
deviations.  Recent improvements with a focus on QCD corrections and
corresponding tools are also discussed.

\end{Abstract}
\vfill

\begin{Presented}
The Fifth Annual Conference\\
 on Large Hadron Collider Physics \\
Shanghai Jiao Tong University, Shanghai, China\\ 
May 15-20, 2017
\end{Presented}
\vfill
\end{titlepage}
\def\thefootnote{\fnsymbol{footnote}}
\setcounter{footnote}{0}
%

\normalsize 


\section{Introduction}

As the heaviest elementary particle known to date, the top quark is special in
many aspects.  It is the only quark that decays semi-weakly, with a short
lifetime, before the hadronization can occur.  It is also the only quark with
order one Yukawa coupling, and for this reason it plays an important role in
the standard model (SM) and in many of its extensions.  Furthermore, the top
quark mass is a crucial parameter related to the vacuum stability, possibly
determining the fate of the universe \cite{Buttazzo:2013uya}. The LHC is a
top-quark factory, providing us with great opportunities to study the top-quark
properties, including its mass, couplings, production rates, decay branching
ratios, etc.  In this talk, we will review some recent theoretical progress
related to the determination of the top-quark couplings.

\section{Parametrization}

The standard model effective field theory (SMEFT) has proved to be an effective
and powerful approach to parametrize our ignorance beyond the SM.  In the
context of searching for deviations from the predicted interactions between the
SM particles, the experimental information on the interactions and possible
deviations can be consistently and systematically interpreted with the SMEFT
approach \cite{Weinberg:1978kz,Buchmuller:1985jz,Leung:1984ni}. The SMEFT
Lagrangian corresponds to that of the SM augmented by higher-dimensional
operators, which respect the symmetries of the SM, and is in particular a
useful approach to identify observables where deviations could be expected in
the top sector \cite{Cao:2007ea,Zhang:2010dr,Degrande:2010kt}. More
importantly, it allows for a global interpretation of measurements coming from
different processes and experiments, which can be consistently incorporated in
one analysis, further enhancing the sensitivity to new physics.

In the top-quark physics, we are interested in dimension-six (dim-6) operators that
enter the most relevant top-quark measurements.  In the so-called Warsaw
basis \cite{Grzadkowski:2010es}, they are given by
\begin{eqnarray}
\noalign{Two-quark operators:}
\nonumber
\begin{array}{lll}
	\hc{\qq{}{u\varphi}{ij}}
	=\bar{q}_i u_j\tilde\varphi\: (\varphi^{\dagger}\varphi)
	,\qquad
	&\qq{1}{\varphi q}{ij}
	=\FDF (\bar{q}_i\gamma^\mu q_j)
	,\qquad
	&\qq{3}{\varphi q}{ij}
	=\FDFI (\bar{q}_i\gamma^\mu\tau^I q_j)
	,\\
	\qq{}{\varphi u}{ij}
	=\FDF (\bar{u}_i\gamma^\mu u_j)
	,\qquad
	&\hc{\qq{}{\varphi ud}{ij}}
	=(\tilde\varphi^\dagger iD_\mu\varphi)
	  (\bar{u}_i\gamma^\mu d_j)
	,\qquad
	&\hc{\qq{}{uW}{ij}}
	=(\bar{q}_i\sigma^{\mu\nu}\tau^Iu_j)\tilde{\varphi}\:g_WW_{\mu\nu}^I
	,\\
	\hc{\qq{}{dW}{33}}
	=(\bar{q}_i\sigma^{\mu\nu}\tau^Id_j){\varphi}\;g_W W_{\mu\nu}^I
	,\qquad
	&\hc{\qq{}{uB}{ij}}
	=(\bar{q}_i\sigma^{\mu\nu} u_j)\tilde{\varphi}\; g_YB_{\mu\nu}
	,\qquad
	&\hc{\qq{}{uG}{ij}}
	=(\bar{q}_i\sigma^{\mu\nu}T^Au_j)\tilde{\varphi}\; g_SG_{\mu\nu}^A
	.
\end{array}
\\
\noalign{Four-quark operators:}
\nonumber
\begin{array}{lll}
	\qq{1}{qq}{ijkl}
	= (\bar q_i \gamma^\mu q_j)(\bar q_k\gamma_\mu q_l)
	,\quad
	&\qq{3}{qq}{ijkl}
	= (\bar q_i \gamma^\mu \tau^I q_j)(\bar q_k\gamma_\mu \tau^I q_l)
	\label{eq:LLLL_2}
	,\quad
	&\qq{1}{qu}{ijkl}
	= (\bar q_i \gamma^\mu q_j)(\bar u_k\gamma_\mu u_l)
	,\\
	\qq{8}{qu}{ijkl}
	= (\bar q_i \gamma^\mu T^A q_j)(\bar u_k\gamma_\mu T^A u_l)
	,\quad
	&\qq{1}{qd}{ijkl}
	= (\bar q_i \gamma^\mu q_j)(\bar d_k\gamma_\mu d_l)
	,\quad
	&\qq{8}{qd}{ijkl}
	= (\bar q_i \gamma^\mu T^A q_j)(\bar d_k\gamma_\mu T^A d_l)
	,\\
	\qq{}{uu}{ijkl}
	=(\bar u_i\gamma^\mu u_j)(\bar u_k\gamma_\mu u_l)
	,\quad
	&\qq{1}{ud}{ijkl}
	=(\bar u_i\gamma^\mu u_j)(\bar d_k\gamma_\mu d_l)
	,\quad
	&\qq{8}{ud}{ijkl}
	=(\bar u_i\gamma^\mu T^A u_j)(\bar d_k\gamma_\mu T^A d_l)
	,\\
	\hc{\qq{1}{quqd}{ijkl}}
	=(\bar q_i u_j)\:\varepsilon\;
	  (\bar q_k d_l)
	,\quad
&	\hc{\qq{8}{quqd}{ijkl}}
	=(\bar q_iT^A u_j)\;\varepsilon\;
	  (\bar q_kT^A d_l)
	  .\label{eq:4f}
\end{array}
\\
\noalign{Two-quark-two-lepton ($2q2l$) operators:}
\nonumber
\begin{array}{lll}
	\qq{1}{lq}{ijkl}
	=(\bar l_j\gamma^\mu l_j)
	  (\bar q_k\gamma^\mu q_l)
	,\qquad
	&\qq{3}{lq}{ijkl}
	=(\bar l_j\gamma^\mu \tau^I l_j)
	  (\bar q_k\gamma^\mu \tau^I q_l)
	,\qquad
	&\qq{}{lu}{ijkl}
	=(\bar l_j\gamma^\mu l_j)
	  (\bar u_k\gamma^\mu u_l)
	,\\
	\qq{}{eq}{ijkl}
	=(\bar e_j\gamma^\mu e_j)
	  (\bar q_k\gamma^\mu q_l)
	,\qquad
	&\qq{}{eu}{ijkl}
	=(\bar e_j\gamma^\mu e_j)
	  (\bar u_k\gamma^\mu u_l)
	,\qquad
	&\hc{\qq{1}{lequ}{ijkl}}
	=(\bar l_i e_j)\;\varepsilon\;
	  (\bar q_k u_l)
	,\\
	\hc{\qq{3}{lequ}{ijkl}}
	=(\bar l_i \sigma^{\mu\nu} e_j)\;\varepsilon\;
	  (\bar q_k \sigma_{\mu\nu} u_l)
	,\qquad
	&\hc{\qq{}{ledq}{ijkl}}
	=(\bar l_i \sigma^{\mu\nu} e_j)
	  (\bar d_k \sigma_{\mu\nu} q_l)
	.
\end{array}
\\
\noalign{Bosonic operators:}
\begin{array}{lll}
	O_{G}=f^{ABC}G^{A\nu}_\mu G^{B\rho}_\nu G^{C\mu}_\rho
	,\qquad\qquad
	&
	O_{HG}=\left( \varphi^\dagger\varphi \right)G_{\mu\nu}^AG^{A\mu\nu}
	.
	&
	\hspace{4cm}
\end{array}
\end{eqnarray}
We have not specified the flavor indices.  For an operator to be relevant
in a SM top process, two of its quark fields need to be of the third generation.
Alternatively, if only one quark field is of the third generation, the operator
can be relevant in the search of top flavor-changing neutral (FCN) interactions.

Two bosonic operators are also included here.  It has been
suggested that the $t\bar t$ production should be used to constrain the
coefficient of $O_{G}$, due to it's non-interference with the SM amplitude
\cite{Cho:1994yu}, even though recently it has been pointed out that the
multijet production process is more efficient \cite{Krauss:2016ely}.  The
$O_{HG}$ operator is included because it formally contributes to the $t\bar t$
process through $pp\to H\to t\bar t$, with a top Yukawa coupling of order one
\cite{Zhang:2010dr}, and also because it is often needed in higher order
calculations, since the chromo-dipole operator $O_{tG}$ mixes into it
\cite{Maltoni:2016yxb}.

\begin{figure}[htb]
\centering
\includegraphics[width=.66\linewidth]{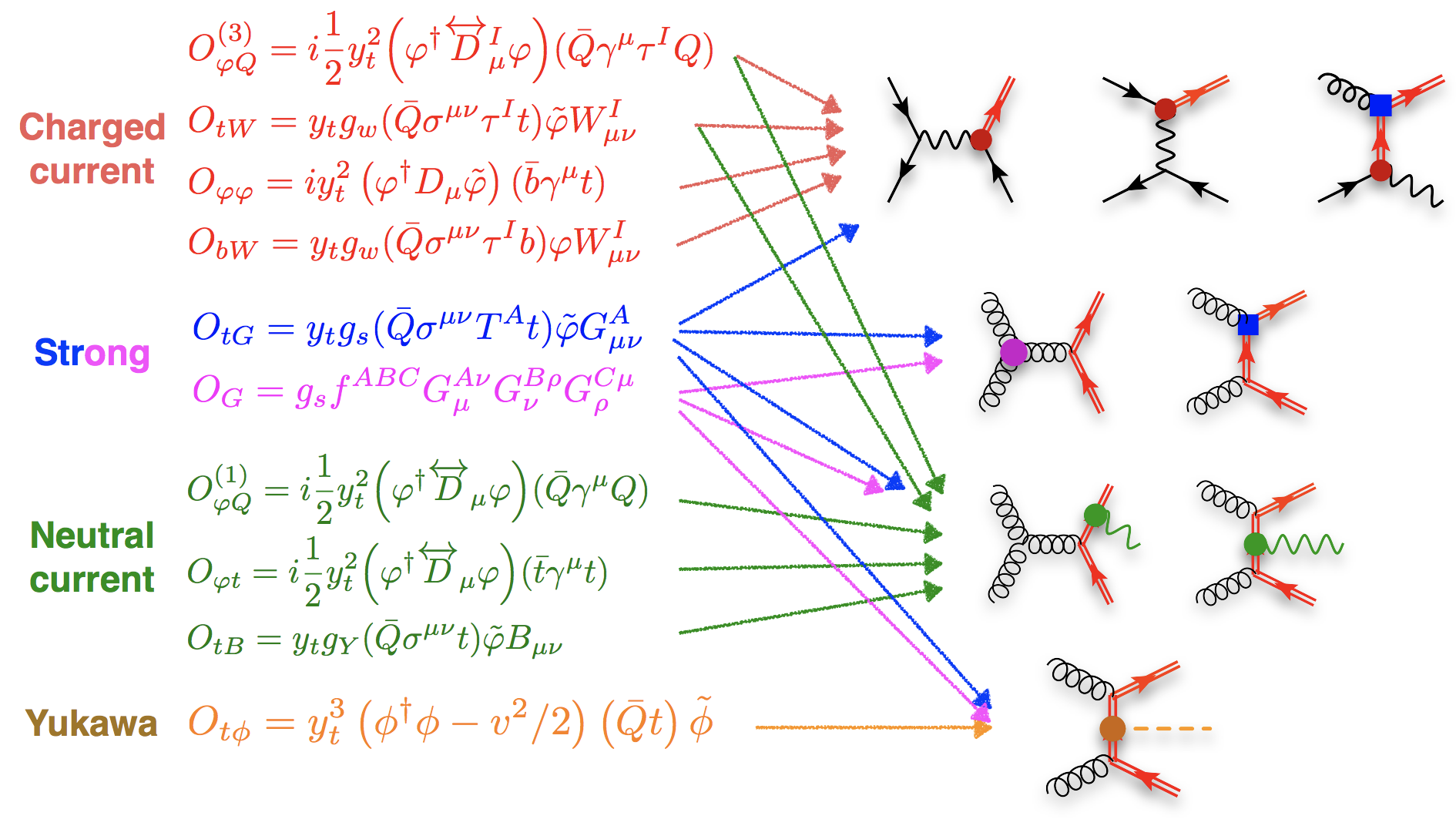}
\caption{ Two-quark operators entering the main production mechanisms,
including single top, $t\bar t$, $t\bar t$+gauge boson and $t\bar t$+Higgs.}
\label{fig:lines}
\end{figure}

In Figure~\ref{fig:lines} we briefly illustrate how the two-quark operators
are related to different top-quark processes. The notation is slightly
different as we have fixed the flavor indices of the quarks.  In particular,
we use capital $Q$ to denote the third-generation $q$ doublet.  In the diagrams,
blobs represent insertions of dim-6 operators, while the double lines represent
top-quark fields.  Intuitively these operators can be roughly
divided into several categories, depending on the most important processes where
they can be probed, including single top, $t\bar t$, $t\bar
t$+gauge boson and $t\bar t$+Higgs.  This allows for an EFT analysis based on
a single process to make sense, and to give meaningful
bounds.  On the other hand, contributions across different categories
exist, and the situation can become more complicated when four-fermion operators
are added.  This motivates the needs for a global analysis, by including all
relevant processes and operators and solving the system, to obtain the most
reliable information from data and also to maximize the sensitivity to all
possible deviations.  
 
\section{Observables}
Global analyses in the literature often follow a bottom-up approach.  They are
based on observables measured under the SM assumptions, which are provided by
experimental collaborations with statistical and systematic uncertainties in
detail, allowing for combination with other measurements.  The set of
observables can be continuously extended as new measurements become available.
The most important advantage of a bottom-up approach is that it can be easily
followed by the theorists, and so any progress on the theory side, such as
improved predictions and combination of new channels, can be immediately
included in real analyses, and the corresponding impacts and improvements can be
investigated in detail.  This is useful for improving our interpretation of the
theory framework and for identifying the most urgent needs.  In the long term it
should also be used as a preparation for a more advanced analysis, which could
directly rely on data (i.e.~without SM observables).

The most obvious observable used in a global fit is the total cross section.
$t\bar t$ and single top cross sections are among the most precisely measured
observables in top measurements, with respectively $\sim 5\%$ and $\sim 10\%$
errors.  Recently at Run II the associated production channels such as $t\bar
t+W/Z$ and single $t+Z$ also reached good precision.  It should be pointed out
that the experimental precision is not directly related to the constraining
power of the measurement.  One example is the four-top production process,
which only has an upper limit of about 4.6 times the SM signal, but its cross
section has a constraining power on $qqtt$ four-fermion operators that is
already comparable with that from the $t\bar t$ measurements, due to an enhanced
sensitivity to $qqtt$ operators \cite{Zhang:2017mls}.  Another example is that
certain processes can benefit from unitarity violation due to electroweak
operators, see a discussion in Ref.~\cite{Dror:2015nkp} about $tW$ scattering.

Differential cross sections play a special role in the SMEFT context.  By naive
power counting, one expects that the dim-6 operator contributions could scale like
$E^2/\Lambda^2$.  A differential measurement can thus isolate the phase space
region with the best sensitivity.  For example, it is well known that in $t\bar t$
production the $m_{tt}$ distribution is powerful, not
only to constrain the four-fermion operators but also to distinguish them from
deviations in other forms \cite{Degrande:2010kt}.  Sensitivity to deviations is
achieved by balancing between small statistical uncertainty and systematic
control for low mass regime but also small new physics-induced deviations, and
the opposite situation for high-mass regime, see Ref.~\cite{Englert:2016aei}
for a more detailed discussion.  

A related class of observables are the asymmetries in $t\bar
t$ production.  They have attracted extensive interest in the past due to
apparent discrepancy between Tevatron data and SM predictions, which then
failed to grow as more data were added.  Taking into account the EW corrections
and next-to-next-to-leading order QCD corrections, the remaining tension
is below 1.5 $\sigma$.  Nevertheless, these observables are useful in
constraining four-fermion operators, as their contributions come from the tree
level.  At the linear order in $C/\Lambda^2$, there are only four independent
degrees of freedom in all four-fermion operators, denoted as $C_{u,d}^{1,2}$.
The cross sections and asymmetries are determined by $C_{u,d}^1+C_{u,d}^2$
and $C_{u,d}^1-C_{u,d}^2$ respectively, and so a useful strategy to bound the
four-fermion operators is to combine cross section measurements with
$A_{FB}$ or $A_C$ measurements, at two different energies (Tevatron and
LHC), which is just enough to constrain all four directions~\cite{Rosello:2015sck}.

The situation can be different if one starts to incorporate
$\mathcal{O}(C^2/\Lambda^4)$ terms. Ref.~\cite{Rosello:2015sck} has pointed out
that within the current experimental bounds, these terms may not be negligible.
In general, if the operator can be inserted once in the amplitude, an
observable will be a quadratic function of the operator coefficients,
\begin{equation}
	\sigma=\sigma_{SM}+\sum_i\frac{C_i\sigma_i}{\Lambda^2}+\sum_{i,j}\frac{C_iC_j\sigma_{ij}}{\Lambda^4}\,.
\end{equation}
The fact that the quadratic term becomes dominant is often interpreted
as the EFT becoming invalid, because naively one could expect that the
dim-8 operators interfere at the same order $\mathcal{O}(\Lambda^{-4})$.
Whether this is really the case is a model-dependent question.
However, we should point out that there are motivations to incorporate
dim-6 squared pieces, while neglecting dim-8 interference pieces.
Suppose the underlying theory is described by one scale $M_*$ and one coupling
$g_*$, so that
\begin{equation}
\frac{C}{\Lambda^2}\sim\frac{g_*^2}{M_*^2}\,. \label{eq:inter}
\end{equation}
The current limits on $C/\Lambda^2$ may imply a large $g_*$ is allowed if $M_*$
is to be kept above the scale of the measurement.  In a process like $q\bar
q\to t\bar t$, one could expect that dim-6 squared terms to dominate, as
they are enhanced by four powers of $g_*$, while the dim-8 interferences to be
subdominant, as they scale at most as $g_*^2$. The EFT expansion is
still valid in this case, see Ref.~\cite{Contino:2016jqw} for more discussions.
A comparison between dim-6 interference and the squared terms is not necessarily related
to the validity of the EFT.

In $t\bar t$ production, if dim-6 squared terms are incorporated, the
$C_{u,d}^{1,2}$ language is no longer valid.  Instead, the full set of 14
four-fermion operators needs to be included
\begin{equation}
	\begin{array}{ll}
{O}^{(8,3)}_{Qq}=\left(\bar Q_L \gamma_\mu T^a \tau^i Q_L\right) \left(\bar q_L \gamma^\mu T^a \tau^i q_L\right)
\qquad&
{O}^{(1,3)}_{Qq}=\left(\bar Q_L \gamma_\mu  \tau^i Q_L\right) \left(\bar q_L \gamma^\mu  \tau^i q_L\right)
\\
{O}^{(8,1)}_{Qq}=\left(\bar Q_L \gamma_\mu T^a Q_L\right) \left(\bar q_L \gamma^\mu T^a q_L\right)
\qquad&
{O}^{(1,1)}_{Qq}=\left(\bar Q_L \gamma_\mu Q_L\right) \left(\bar q_L \gamma^\mu  q_L\right)
\\
{O}^{(8)}_{td}=\left(\bar t_R \gamma_\mu T^a  t_R\right) \left(\bar d_R \gamma^\mu T^a d_R\right)
\qquad&
{O}^{(1)}_{td}=\left(\bar t_R \gamma_\mu  t_R\right) \left(\bar d_R \gamma^\mu  d_R\right)
\\
{O}^{(8)}_{tu}=\left(\bar t_R \gamma_\mu T^a  t_R\right) \left(\bar u_R \gamma^\mu T^a u_R\right)
\qquad&
{O}^{(1)}_{tu}=\left(\bar t_R \gamma_\mu  t_R\right) \left(\bar u_R \gamma^\mu  u_R\right)
\\
{O}^{(8)}_{tq}=\left(\bar t_R \gamma_\mu T^a  t_R\right) \left(\bar q_L \gamma^\mu T^a q_L\right)
\qquad&
{O}^{(1)}_{tq}=\left(\bar t_R \gamma_\mu  t_R\right) \left(\bar q_L \gamma^\mu  q_L\right)
\\
{O}^{(8)}_{Qd}=\left(\bar Q_L \gamma_\mu T^a Q_L \right)\left(\bar d_R \gamma^\mu T^a d_R\right)
\qquad&
{O}^{(1)}_{Qd}=\left(\bar Q_L \gamma_\mu Q_L \right)\left(\bar d_R \gamma^\mu  d_R\right)
\\
{O}^{(8)}_{Qu}=\left(\bar Q_L \gamma_\mu T^a Q_L \right)\left(\bar u_R \gamma^\mu T^a u_R\right)
\qquad&
{O}^{(1)}_{Qu}=\left(\bar Q_L \gamma_\mu Q_L \right)\left(\bar u_R \gamma^\mu  u_R\right)
\end{array}
\end{equation}
assuming a U(2)$^{3(q,u,d)}$ flavor symmetry on the first two generations.
They represent 14 independent degrees of freedom in $t\bar t$ production, and
can be derived from Eq.~(\ref{eq:4f}), by assigning appropriate flavor indices.

\begin{figure}[htb]
	\begin{center}
		\includegraphics[width=.98\linewidth]{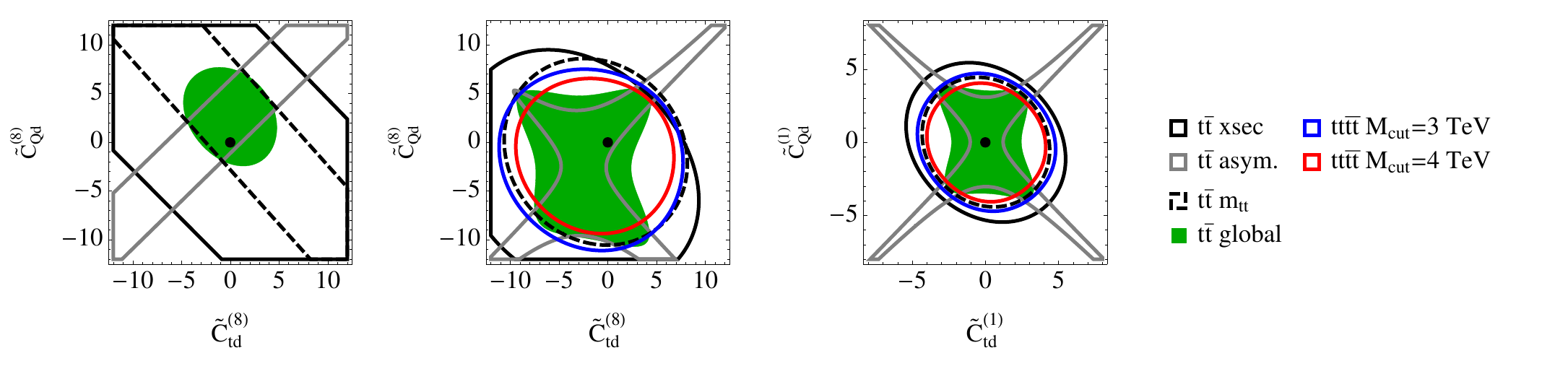}
		\\
		(a)\hspace{3.45cm}(b)\hspace{3.45cm}(c)\hspace*{3.6cm}
	\end{center}
	\vspace{-8pt}
	\caption{Constraints from $t\bar t$ cross section measurements
		(black solid), asymmetry measurements (gray), differential
		$m_{tt}$ distribution (black dashed), and four-top projection
		(blue and red) on selected four-fermion operators.  Green
		shaded area corresponds to the combined result from $t\bar t$
		measurements (see Ref.~\cite{Zhang:2017mls} for details).
		Operators in (a), (b) interfere with the SM amplitude, while
		those in (c) do not due to different color structures.  Dim-6
		squared and higher power terms are neglected in (a), but kept
		in (b), (c).  }
	\label{fig:tt}
\end{figure}

For illustration purpose, in Figure~\ref{fig:tt} we give the 95\% CL
constraints from
cross section measurements, asymmetry measurements, a differential $m_{tt}$
distribution measurement, as well as bounds from four-top production.  In
Figure~\ref{fig:tt} (a) we show constraints on two color-octet operators, where
only the dim-6 linear terms are kept.  The constraints from cross section and
$m_{tt}$ are orthogonal to those from asymmetry measurements, as they are
determined by $C_d^1\pm C_d^2$ respectively.  In Figure~\ref{fig:tt} (b) we
consider the same operators, but without truncating the higher power terms of
$C/\Lambda^2$.  One can see that the allowed region from each experiment
becomes quite different, as the dominant contributions become $(C_d^1)^2\pm
(C_d^2)^2$.  Interestingly, beyond the linear term, the four-top production
cross section can place competitive constraints.  In the plot we show the
projections for $300$ fb$^{-1}$, with analysis cuts, $M_{cut}=3$ and 4 TeV,
applied on the center of mass energy to ensure the validity of the EFT, see
Ref.~\cite{Zhang:2017mls} for more details.  Finally, in Figure~\ref{fig:tt}
(c) we show similar constraints for two color singlet operators.  These
operators do not interfere with the SM at the dim-6 linear order, so no
constraints can be obtained at that order.  However, we can see that by
including the dim-6 squared terms, the resulting constraints are even tighter
than those of the color octet operators which interfere, due to a larger color
factor.

Finally, the decay process also provides useful information.  The most
constrained observable is the $W$-helicity fraction. It is sensitive to the
operator $O_{uW}^{(33)}$, but unlike the single-top cross
section, it is virtually independent of the other operators that enter the
$Wtb$ vertex, and thus providing complementary information.  It should be noted
that the $W$-helicity fraction is only a pseudo-observable, as the $W$ boson
does not exist if the top decays through a four-fermion operator, which limits
the applicability of this observable in a global fit.  The total width, on the
other hand, has not been measured to a very accurate level.  This can be an
issue because in principle one only measures the total cross section times the
branching ratio, so the width could affect all cross section measurements if
the theory framework used allows for exotic decay channels.  In this respect it
is useful to consider new methods that could improve the width measurement.
For example, in Ref.~\cite{Giardino:2017hva} it has been proposed to use the
$b$-charge identification as a way to remove the background and to improve the
precision of the width measurement at high integrated luminosity.

A global fit that takes into account most of the existing measurements has
been performed in Refs.~\cite{Buckley:2015nca,Buckley:2015lku}. It is perhaps
not surprising that no significant deviation from the SM has been found.
Constraints are in general not very tight, in the sense that if one interprets
the results using Eq.~(\ref{eq:inter}), and keeps $M_*$ above the energy of the
measurement, then only a small window of the parameter space can be excluded if
one requires the underlying theory to be perturbative \cite{Englert:2016aei}.
Still, further improvements are expected at high integrated luminosity, as many
differential distributions are still dominated by statistics, in particular in
high mass regions where the EFT sensitivity is the largest.

\section{NLO phenomenology}
\label{sec}
As we await for more results from experiments, the theory approach continues to
improve.  A lot of studies have appeared in the literature, some focusing on the
matching of SMEFT to specific models (see, e.g., Ref.~\cite{Zhang:2016pja} and
references therein), while others aiming at improving the precision of the theory
prediction itself.  We focus on the latter.

Given that the expectations from LHC Run-II on the attainable precision of the
top-quark measurements are high, next-to-leading order (NLO) predictions for
top-quark production channels are becoming relevant. Due to the large number of
the relevant operators, we rely on many different observables predicted by each
operator to effectively constrain every possible deviations from the SM.  It is
then important to know if and how these observables are modified by QCD
corrections, in particular given that the top-quark is a colored particle. 

Recently, NLO predictions for the SMEFT, matched with parton shower simulation,
are becoming available in the {\sc MadGraph5\_aMC@NLO} framework
\cite{Alwall:2014hca}, based on an automatic approach to NLO QCD calculation
interfaced with shower via the {\sc MC@NLO} method \cite{Frixione:2002ik}.  The
dim-6 Lagrangian can be implemented with the help of a series of
packages, including {\sc FeynRules} and {\sc NLOCT} \cite{Alloul:2013bka,
Degrande:2014vpa}.
A model in the Universal {\sc FeynRules} Output format \cite{Degrande:2011ua}
can be built, allowing for simulating a variety of processes at NLO in QCD. We
briefly review the recent progresses in this direction. The interested readers
may find more details in
Refs.~\cite{Degrande:2014tta,Franzosi:2015osa,Zhang:2016omx,Bylund:2016phk,Maltoni:2016yxb}.
With these new implementations, top-quark analyses based on SMEFT are being
promoted systematically to NLO in QCD.

\begin{figure}[ht]
 \begin{minipage}{\textwidth}
  \begin{minipage}[b]{0.44\textwidth}
	  \centering
	  \includegraphics[width=.9\linewidth,height=.66\linewidth]{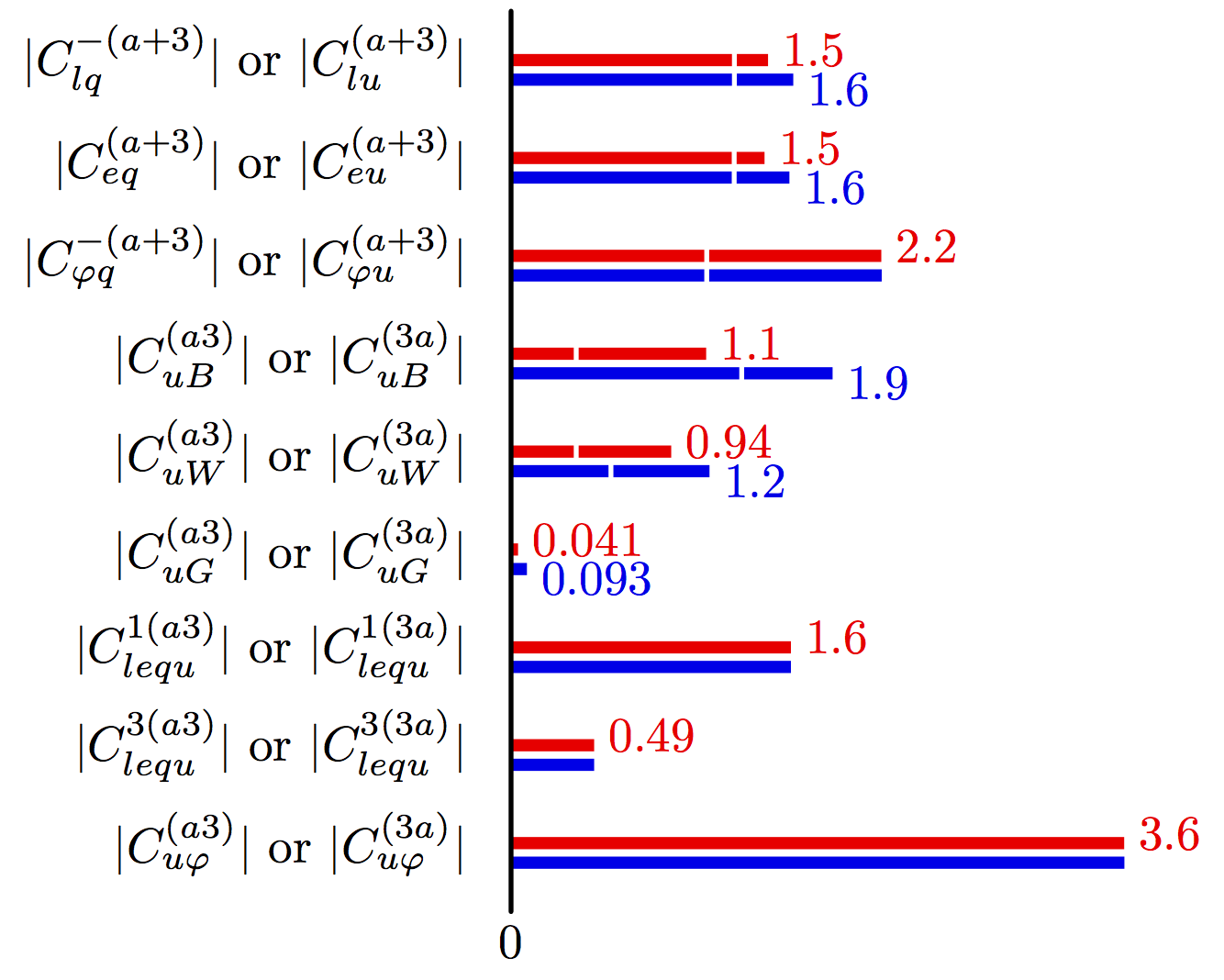}
		\captionof{figure}{Global constraints on FCN interactions,
			assuming $\Lambda=1$ TeV. The red (blue) allowed
			regions apply for a = 1 (2). A white mark indicates
			the individual limit.  See Ref.~\cite{Durieux:2014xla}
			for more details.
			\label{fig:fcnc}}
  \end{minipage}
  \begin{minipage}[b]{0.56\textwidth}
	  \centering
		\includegraphics[width=.96\linewidth]{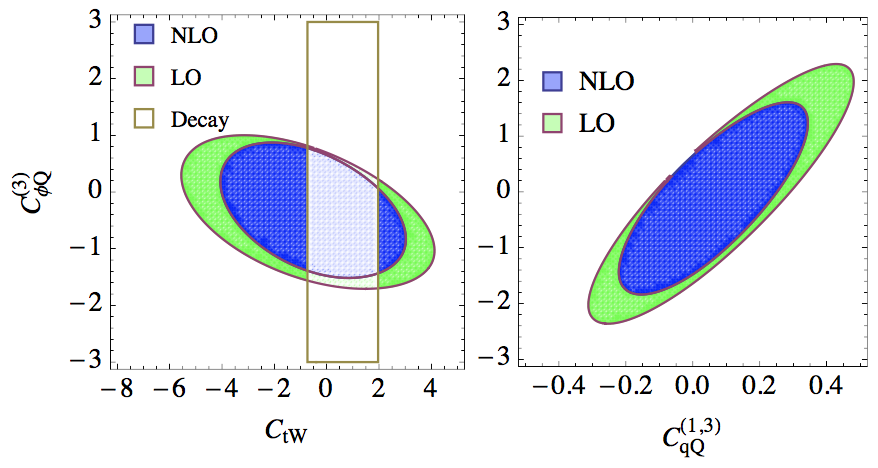}
		\captionof{figure}{95\% limits from single-top measurements,
			with LO or NLO predictions in EFT, assuming $\Lambda=1$
			TeV, see Ref.~\cite{Zhang:2016omx} for more details.
			\label{fig:singlet}}
	\end{minipage}
    \end{minipage}
\end{figure}

\textbf{\textit{Single top FCN production.}}
The first step along this direction was made by Ref.~\cite{Degrande:2014tta},
where the dim-6 operators that give rise to FCN interactions have been
implemented in the framework described above. Single top production processes
in association with a neutral gauge boson or a Higgs boson through FCN
interactions have been computed at NLO in QCD, and corrections are found to be
large.  These results are relevant for phenomenology studies.  In
Ref.~\cite{Durieux:2014xla}, we have performed a simple fit for all FCN interaction
operators (including $2q2l$ ones), to demonstrate that in principle the
framework allows for global fits to be performed at the NLO accuracy, given that
corresponding tools are available.  Some results are given in Figure~\ref{fig:fcnc}.

\textbf{\textit{Top-pair production.}}
The chromo-dipole operator for the top quark, $O_{tG}\equiv y_tO_{uG}^{(33)}$,
can be constrained by top-pair production.  Assuming real operator coefficient,
this calculation has been carried out at NLO in Ref.~\cite{Franzosi:2015osa}.
The $K$-factors for the total cross sections are found to be $1.1$, $1.4$, and
$1.5$ respectively for Tevatron, LHC 8 TeV, and LHC 13/14 TeV. As a result, the
current limits on the chromo-magnetic dipole moment of the top quark from
direct measurements can be improved by roughly the same factors.  

\textbf{\textit{Single top production.}}
Single top production has been computed in all three channels ($t$-channel,
$s$-channel, and $tW$ associated production channel) at NLO in QCD, with
operators $O_{\phi Q}^{(3)}\equiv y_t^2/2O_{\varphi q}^{3(33)}$, $O_{tW}\equiv
y_t O_{uW}^{(33)}$, and $O_{qQ}^{(1,3)}\equiv O_{Qq}^{(1,3)}$
\cite{Zhang:2016omx}.  Inclusive $K$-factors typically range from
$\sim10\%$ to $\sim50\%$.  Scale uncertainties are significantly reduced. A
three-operator fit based on cross section measurements have been performed.
Results are shown in Figure~\ref{fig:singlet}, illustrating the improvements
due to including QCD corrections.

\textbf{\textit{Top-pair production in association with a gauge boson.}}
At the LHC, the neutral couplings $ttZ$ and $tt\gamma$ can be probed
by associated production of a top-quark pair with a neutral gauge boson
$Z/\gamma$.  The relevant operators are
$O_{\phi Q}^{(1)}\equiv y_t^2/2 O_{\varphi Q}^{33}$,
$O_{\phi t}\equiv y_t^2/2 O_{\varphi u}^{33}$,
$O_{tB}\equiv y_tO_{uB}^{33}$,
as well as $O_{tG}$, $O_{\phi Q}^{(3)}$, and $O_{tW}$.
The corresponding NLO predictions are given in Ref.~\cite{Bylund:2016phk}.
By studying the differential distributions, we find that the differential
$K$-factor of the SM and that of the operator contribution can be quite
different, therefore using the SM $K$-factor to rescale the operator
contributions may not be a good approximation.  The authors of
Refs.~\cite{Rontsch:2014cca,Rontsch:2015una} have considered the same process
taking into account the decay of the top quarks.  Constraints from both high
luminosity LHC and ILC are obtained, and significant improvements due to QCD
corrections are found, as a consequence of the reduced scale uncertainty and
the larger event rate due to a positive perturbative correction.

\newcommand{\xs}[7]{$#1^{+#2+#6+#4}_{-#3-#7-#5}$}
\begin{figure}[ht]
 \begin{minipage}{\textwidth}
  \hfill
  \begin{minipage}[b]{0.38\textwidth}
	  \small
	  \centering
		\includegraphics[width=.95\linewidth]{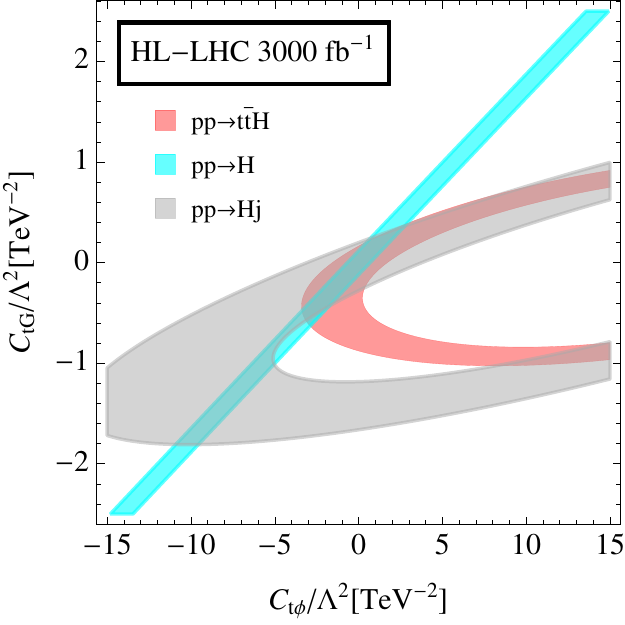}
		\captionof{figure}{Projected constraints from $t\bar tH$, $H$,
			and $H+j$ at the LHC, 3000 fb$^{-1}$.\label{fig:tth1}}
	\end{minipage}
  \hfill
  \begin{minipage}[b]{0.5\textwidth}
	  \small
	  \centering
		\includegraphics[width=.9\linewidth]{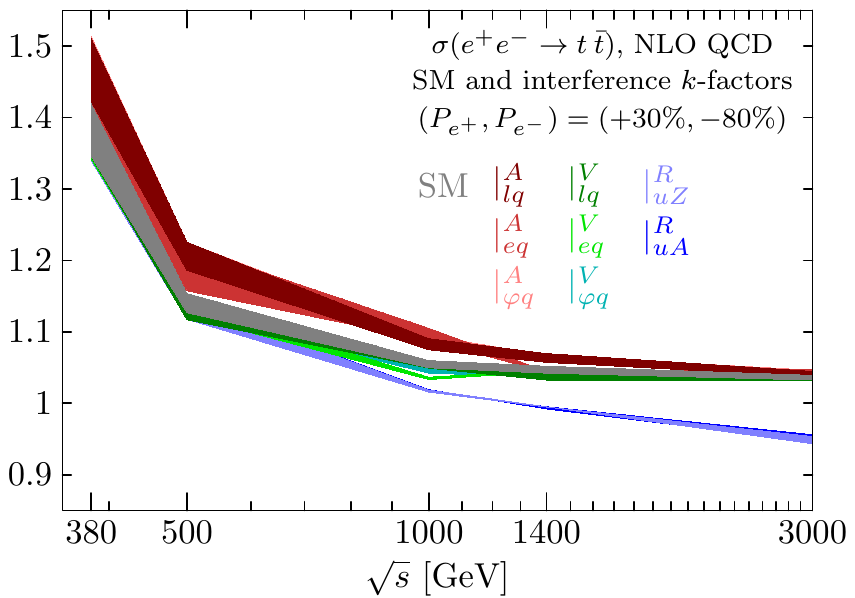}
		\captionof{figure}{The QCD-NLO $K$-factors for the
			$e^+e^-\to t\bar t$ cross section, as a function of the
			centre-of-mass energy. The thickness of the bands
			covers QCD renormalization scale variation between $m_t/2$
			and $2m_t$.\label{fig:eettk}}
  \end{minipage}
  \hfill{}
    \end{minipage}
\end{figure}

\textbf{\textit{Top-pair production in association with a Higgs boson.}}
The LHC provides the first chance to directly measure the interactions between
the top quark and the Higgs boson through the associated production of a Higgs with
$t\bar t$.  In Ref.~\cite{Maltoni:2016yxb}, this process has been computed
at NLO including three operators: the chromo-dipole operator 
$O_{tG}$, the Yukawa operator
$O_{t\phi}\equiv y_t^3 O_{u\varphi}^{(33)}$, and the Higgs-gluon operator
$O_{\phi G}\equiv y_t^2 O_{HG}$. 
The QCD mixing of these three operators goes in the direction of increasing
number of Higgs fields, i.e.~$O_{tG}$ mixes into $O_{\phi G}$, and both of them
mix into $O_{t \phi}$, but not the other way around.
Based on the full NLO results, a combined fit using $t\bar tH$, $H$, and $H+j$
production has been performed to derive the current constraints and future
projections (see Figure~\ref{fig:tth1}).  The results imply
that the Higgs measurements have started to become sensitive to the
chromo-dipole coupling of the top, and so in future analyses the Higgs data
will play a role in the study of top properties.

\textbf{\textit{Top-pair production at $e^+e^-$ colliders.}}
This process can be computed at NLO in QCD with not only the two-quark
operators, but also the $2q2l$ operators, in both the $t\bar t$ and
the $bW^+\bar bW^-$ final states.  The global constraints from measurements at
the future lepton colliders are under investigation \cite{progress}.  In
Figure~\ref{fig:eettk} we illustrate the sizes of QCD corrections to the
interference terms from different operators.

\section{Outlook}
So far we have only considered bottom-up analyses. Ideally, one could improve
the experimental
sensitivity by making use of the accurate SMEFT predictions and designing
optimized experimental strategies in a top-down way.  This is well established
and widely used for concrete new physics models, but have been rarely
considered in the standard SMEFT context.  Given that the aim is to maximize
the discovery potential of the LHC by combining state-of-the-art theoretical
techniques and direct access to experimental data with full details, it is
important to have all the theory set up ready before starting such an effort.
The recent theoretical developments are paving the way towards this goal:
we have discussed improvements on the precision and accuracy of theoretical
predictions and corresponding tools, while another relevant issue, namely
the optimization of the operator basis for top physics, is also under investigation
(see \cite{basis} for some preliminary discussions).  As a first step towards the
final goal, preliminary studies on $t\bar t$ production based on a
matrix-element method have shown improved sensitivity with respect to SM
observable-based analysis \cite{talk}.  We hope to see more progresses along
this direction in the future to further push our reach in determining the
top-quark properties.

\Acknowledgements
I would like to thank C.~Degrande, O.~B.~Bylund, G.~Durieux, D.~B.~Franzosi,
F.~Maltoni, M.~P.~Rosello, I.~Tsinikos, M.~Vos, E.~Vryonidou, and J.~Wang for
collaboration on many works mentioned in section \ref{sec}.

\end{document}

%% file: econfmacros.tex
\def\beq{\begin{equation}}
\def\eeq#1{\label{#1}\end{equation}}
\def\eeqn{\end{equation}}

\def\beqa{\begin{eqnarray}}
\def\eeqa#1{\label{#1}\end{eqnarray}}
\def\eeqan{\end{eqnarray}}

\let\bar=\overbar

\def\hc{{\mbox{\rm h.c.}}}

\def\Dslash{\not{\hbox{\kern-4pt $D$}}}
\def\dslash{\not{\hbox{\kern-2pt $\del$}}}

\def\msb{{\bar{\ssstyle M \kern -1pt S}}}

